\documentclass[11pt, a4paper]{article}

\title{Axion}

\usepackage[usenames, dvipsnames]{color}
\usepackage[T1]{fontenc}
\usepackage[latin1]{inputenc}
\usepackage{lmodern}
\usepackage{amsmath}
\usepackage[pdftex]{graphicx}
\usepackage{lastpage}
\usepackage{amssymb} 
\usepackage{hyperref}

\usepackage{cite}

\usepackage{titling}
\usepackage{authblk}

\usepackage{yfonts}

\usepackage{esvect}
\renewcommand*\vec{\vv}

\usepackage{xspace}

\newcommand{\ex}{\text{e}}
   
   \usepackage{hyphenat}   
   
\usepackage{slashed}
\usepackage{braket}
\usepackage{simplewick}
\usepackage{bbm}

\usepackage{mathtools}

\usepackage{xifthen}
\newcommand*\diff{\mathrm{d}} 
\newcommand*\ldiff[2][]{ \ifthenelse{\isempty{#1}}{ \diff #2}{\diff^#1#2} \,} 
\let\limitint\int 
\renewcommand{\int}{\limitint \!} 

\makeatletter
\renewcommand\section{\@startsection {section}{1}{\z@}%
	{-3.5ex \@plus -1ex \@minus -.2ex}%
	{2.3ex \@plus.2ex}%
	{\normalfont\large\bfseries}}
\makeatother

\title{Classicality and Quantum Break-Time for Cosmic Axions\\[1.5\baselineskip]}

\author[1,2,3]{Gia Dvali}
\author[1,2]{Sebastian Zell\thanks{sebastian.zell@campus.lmu.de}}
\affil[1]{Arnold Sommerfeld Center, Ludwig-Maximilians-Universit\"at, \mbox{Theresienstraße 37, 80333 M\"unchen, Germany}}
\affil[2]{Max-Planck-Institut für Physik, F\"ohringer Ring 6, 80805 M\"unchen, Germany}
\affil[3]{Center for Cosmology and Particle Physics, Department of Physics,\mbox{ New York University, 4 Washington Place, New York, NY 10003, USA}}

\begin{document}

\allowdisplaybreaks

\maketitle

\begin{abstract}

We investigate the length of the period of validity of a classical description 
for the cosmic axion field. 
 To this end, we 
 first show that we can understand the oscillating axion solution as expectation value over an underlying coherent quantum state. Once we include self-interaction of the axion, the quantum state evolves so that the expectation value over it 
 starts to deviate from the classical solution. The time-scale of this process defines the quantum break-time. For the hypothetical dark matter axion field in our Universe, we show that quantum break-time exceeds the age of the Universe by many orders of magnitude. This conclusion is independent of specific properties of the axion model. Thus, experimental searches based on the classical approximation of the oscillating cosmic axion field are fully justified. Additionally, we point out that the distinction of classical nonlinearities and true quantum effects is crucial for calculating the quantum break-time in any system.  Our analysis can also be applied to other types of  dark matter that are described as classical fluids in the mean field approximation.

\end{abstract}

\clearpage
\tableofcontents

\section{Quantum Breakdown of Classicality }

Classicality is an approximate notion, applicable to macroscopic systems with a large number of quantum constituents. The dynamics of such systems is usually well-described by classical equations of motion. However, since the underlying dynamics is quantum, it is legitimate to ask how fast a given macroscopic system can deviate from the classical evolution, due to quantum effects.  This question was addressed in \cite{nico} using a 
prototype many-body system of $N$ interacting bosons in a box with 
periodic boundary conditions.  The bosons experienced a simple attractive four-point interaction controlled by a quantum coupling constant $\alpha$.  
The usefulness of this model lies in the fact that, on the one hand, it is complex enough to exhibit both regimes of interest, i.e., approximately classical versus fully quantum.  On the other hand, it is simple enough to explicitly track the exact quantum evolution and determine when the classical description breaks down.  Following \cite{nico}, we shall refer to this phenomenon as {\it quantum breaking} and to the corresponding time-scale as {\it quantum break-time}.
In \cite{nico} it was observed  that the 
quantum break-time was the shortest (scaling as $\sim \ln(N)$) for states that exhibited a classical instability with large Lyapunov exponent and a strong collective coupling  $\lambda \equiv \alpha N$.  
At the same time, for the states with weak collective coupling 
and/or no classical instability, the quantum breaking time is macroscopically-long and scales as power-law in $N$. For describing  the time-evolution of such states, the classical approximation therefore remains valid during a time that is power-law sensitive to the occupation number of bosons.   

Since \cite{nico}, the question of the duration of a classical description in quantum many-body systems  has been studied for  various multi-boson states of interacting spin-2 and spin-0 particles \cite{compositenessDS, kuhnel, sikivieCoherent, hertzbergClassical, berezhiani, us}. In the present paper, we shall adopt the approach of \cite{us},  where the concept of quantum break-time was applied to the field theoretic system of an oscillating nonlinear scalar field and certain universal relations  among different time-scales were derived. 
As we shall see, these relations allow to clearly separate the effects  of quantum  interactions from those of  classical nonlinearities. This analysis confirms that for weak collective coupling and in the absence of classical  Lyapunov exponents, the quantum breaking-time is macroscopically long.  In particular, our goal in the present paper is to apply the above analysis to a cosmic axion field and to understand the domains of validity of its classical description.

\section{Importance of a Classical Description of Dark Matter Axions}

The axion \cite{weinberg, wilczek} is a well-known hypothetical particle which is predicted by the Peccei-Quinn (PQ) solution \cite{pecceiquinn} to the strong CP problem. It is a pseudo-Goldstone boson of spontaneously-broken global chiral PQ symmetry. The explicit breaking of this symmetry by the chiral anomaly through non-perturbative QCD effects results in a non-zero mass of the axion. One of the beauties of axion physics  is that its low energy dynamics is extremely constrained due to the  Goldstone nature and the power of anomaly. 
The mass $m_a$  and the decay constant $f_a$ of the axion field are related via a non-perturbative scale $\Lambda$: $m_a \,  = \, {\Lambda^2 \over f_a}$.  Note that in any theory in which the sole source of the PQ symmetry breaking is the QCD anomaly, the scale $\Lambda$ is entirely determined by the non-perturbative QCD sector and 
the low-energy parameters of the Standard Model 
(such as the Yukawa coupling constants of quarks) and is insensitive to the precise embedding of the axion into a high-energy theory, i.e., low energy axion physics is insensitive with respect to UV-completion.
 
Phenomenological constraints put a lower bound on the scale $f_a$ approximately around $10^{9}\,$GeV (see e.g., \cite{review}).
This speaks in favor of so-called invisible axion models \cite{invisible}, in which the PQ symmetry can be broken at a very high scale around $f_a \gtrsim 10^{9}\,$GeV.  Such a weak coupling implies that the axion is essentially stable on cosmological scales. This fact makes the axion a very interesting dark matter candidate. 
In this scenario,  the role of dark matter energy is played by the energy of coherent oscillations of the axion field.  Of course, the current energy of the axion field 
 depends on the cosmological epoch of the onset of axion oscillations  
as well as on its initial amplitude $A_{\text{in}}$. 
 Some conservative estimates are based on the assumptions that the axion oscillations first started in the epoch of thermal phase transition of  QCD and  
 with the maximal  initial amplitude $A_{\text{in}} \sim f_a$. 
 This gives the famous cosmological upper bound: $f_a \lesssim 10^{12}\,$GeV \cite{cosmicbound}. We must notice that there exist loopholes \cite{gia} 
 which soften this upper bound and allow for much higher values of $f_a$.     
 For the present study, however, this change is unimportant  
and we can safely assume $f_a$ to be below its conservative upper bound, $f_a \lesssim 10^{12}\,$GeV.    
 
The search for the dark matter axion has been an active field of research since no signs of it have been found so far (see e.g., \cite{experiments} for current experimental efforts). 
 Several of the proposed experiments heavily rely on the approximation of the gas of axions by a coherently oscillating classical scalar field $a(t)$.  There has been a recent discussion of the axion field on the quantum level (see e.g., \cite{sikivieFirst, sikivieDetailed, sikivieAngular}). Although the main motivation there was astrophysical, it was also suggested that quantum effects can significantly correct the classical description of axions. 
Obviously, it is important to clarify this issue both from a fundamental as well as an experimental point of view.   
 In doing so, we will not discuss the astrophysical consequences proposed in  \cite{sikivieFirst, sikivieDetailed, sikivieAngular}. Instead, we will only be concerned with the general question if the classical description as oscillating scalar field is valid for axions.

In the present paper, we will therefore calculate a  lower bound on the quantum break-time of the cosmic axion field, i.e., the minimal required  time-scale before the true quantum evolution of a multi-axion quantum state can depart from its classical mean field description. As said above, the study of the quantum break-time phenomenon  for an oscillating scalar field was given in \cite{us}, and we shall apply this analysis to the axion field. 
Using these quantum field-theoretic arguments, we will show that if for some initial time the approximation of an axion gas by a classical field is good, it remains good -- with an extraordinary accuracy  -- for time-scales exceeding the current age of the Universe by many orders of magnitude.  Hence, experimental searches relying on the classical field approximation are safe for all practical purposes.  

We start in section \ref{sec:scaling} by discussing the scaling of the quantum break-time in a generic system. We turn to the axion field in section \ref{sec:classical} and review some of its classical properties. Subsequently, we restrict ourselves to a simplified setup which possesses an explicit solution. In section \ref{sec:quantum}, we will show that we can reproduce this simple classical solution as expectation value over a coherent state of axions. Once we include the selfcoupling of axions, this coherent state starts to evolve so that the expectation value over it departs from the classical solution. The time-scale of this process is the sought-after quantum break-time and thereby the main result of our paper. Afterwards, we show in section \ref{sec:simplifications} that the approximations we used to find our simple classical solution do not alter our conclusion. In particular, we show that Hubble damping has no influence on the quantum break-time.  We supplement this discussion by elaborating on the collective coupling of the axion in section \ref{sec:collective}. While it is weak for the hypothetical dark matter axion in our Universe, we demonstrate that it would not lead to an arbitrarily short break-time even if it were infinitely strong. We conclude in section \ref{sec:otherWork} by discussing in more detail the relationship of our work to the results presented in \cite{sikivieFirst, sikivieDetailed, sikivieAngular}. In particular, we argue that the effects observed there also occur in a classical mean field description and do not lead to quantum breaking.

\section{General $\hbar$-Scaling of Time-Scales} 
\label{sec:scaling}
 
 Before going into axion specifics, it is useful to discuss some general features  of classicality for a generic multi-particle quantum system. A key characteristic of any quantum field-theoretic system is the strength of interaction between its quanta, which can be parameterized by a dimensionless quantity $\alpha$. In the quantum language, $\alpha$ controls the magnitude of scattering amplitudes.  Typically, it is convenient to use  $2 \rightarrow 2$-scattering as a reference point. 
 Of course, the system may possess more than one type of interaction, and correspondingly more than one type of $\alpha$. However, for purposes of this discussion a single $\alpha$ is sufficient.   
  
  We can always normalize fields in such a way that $\alpha \ll 1$ corresponds to a weak-coupling domain, in which a perturbative expansion in powers of $\alpha$ can be performed. Correspondingly, $\alpha >  1$ describes a {\it strong coupling} regime, for which perturbation theory in $\alpha$ breaks down.  In our analysis we shall restrict ourselves to systems with weak coupling since the axion satisfies this property.    
   Even for  $\alpha \ll 1$, however, the system can become strongly interacting in a {\it collective} sense.  This can happen if the system is  put in  a state in which the occupation number of interacting quanta $N$ is large enough. In that case, the strength of interaction is determined by the collective coupling
   \begin{equation}
   \lambda \equiv \alpha N \,.
   \label{collectiveGeneral}
   \end{equation}
   The regimes of interest can be then split according to whether it is weak ($\lambda < 1$), strong ($\lambda > 1$) or critical ($\lambda = 1$).   

Let us investigate how the various quantities scale when we take  
the classical limit.  This limit can be defined in several equivalent ways.  One possibility is to take $\hbar \rightarrow 0$, while keeping all the classically-measurable  expectation values fixed, i.e., all the parameters in the classical Lagrangian are kept finite.  In particular, the collective coupling $\lambda$ is a classical quantity  since it characterizes the strength of classical nonlinearities. Therefore, it is independent of $\hbar$ and stays finite in the classical limit. Since the quantum coupling vanishes, $\alpha \rightarrow 0$, for $\hbar \rightarrow 0$, this implies that  in the classical limit we have $N \rightarrow \infty$. Therefore, states which behave approximately-classically are characterized by a large occupation number of quanta $N$.  In particular, this is true for the coherently oscillating axion field.   
 
 Keeping in mind that we are at large $N$, small $\alpha$ and some fixed $\lambda$, we can now perform some dimensional analysis. We assume that  the system is well-described classically at some initial time 
 $t=0$ and we wish to estimate how long it will take for the classical description to break down. Obviously, this time-scale should satisfy the scaling property of becoming infinite in the classical limit $\hbar \rightarrow 0$. Then assuming a simple analytic dependence, the quantum break-time should scale to leading order as 
\begin{equation} 
t_{\text{q}}  =  ({\rm fixed~classical~quantity  } ) \times  N^{\beta}  =  ({\rm fixed~classical~quantity}) \times
  \alpha^{-\beta} \,,
\label{scaling}
\end{equation} 
 where $\beta > 0$ is an integer and we used that $\lambda = \alpha N$ is a fixed constant.  Already for $\beta = 1$, this time-scale is very large for macroscopically-occupied weakly-interacting systems. As we shall see, this is exactly the case for the axion field.  
 
  For completeness, we want to point out that there is another possible functional dependence which  fulfills a simple scaling behavior in the classical limit and which requires special attention: 
  \begin{equation} 
t_{\text{q}}  =   ({\rm fixed~classical~quantity} ) \times  \ln(N)  =  ({\rm fixed~classical~quantity} ) \times
  \ln(\alpha^{-1}) \,.
\label{scalingLog}
\end{equation} 
 Such a scaling cannot be excluded on the basis of general dimensional analysis.  In fact, it was explicitly shown in \cite{nico} that it does take place, but under the following conditions:\\ 
  1) the system must be in an overcritical state, i.e., in a state with  $\lambda > 1$;
 and  \\ 
 2) the system in this state must exhibit a classical instability, i.e., a Lyapunov exponent which is independent of $\hbar$.\\
 Under such conditions, the quantum break-time was found to be given by 
   \begin{equation} 
t_{\text{q}}  =   \Omega^{-1} \times  \ln(N) \, , 
\label{scalingLog1}
\end{equation} 
 where $\Omega$ is the Lyapunov exponent.

   As we shall discuss in detail in section \ref{sec:collective}, the axion field of cosmological interest  does not satisfy either of the above conditions:  1) it is well under-critical,  $\lambda  \ll  1$; 
 and 2) no classical Lyapunov exponent exists. 
     Correspondingly, when modeling the axion field as a many-body quantum system, one should remember to keep $\lambda$ small.  
   
 Moreover, as we shall explain later in the paper, 
 once the underlying Goldstone nature of the axion field is taken into account, the validity of the effective field theory description of the axion precludes  taking the limit $\lambda \gg 1$.  Thus, a short quantum break-time cannot be achieved even at such an expense. 
 Of course, such a regime is not of any obvious 
 phenomenological interest since, as said above, the realistic    
cosmic axion field corresponds to $\lambda \ll 1$. 
Nevertheless,  it is illuminating to clarify this point.

\section{Classical Solutions}
\label{sec:classical}

\subsubsection*{General Properties}
We are now ready to specialize to the axion case. In order to derive a lower bound on the time-scale over which the classical evolution can be a good approximation for an oscillating axion field, the precise form of its potential $V(a/f_a)$ is not so important.  For our 
analysis, it suffices to assume that it is a periodic function of  $a/f_a$.  Only later, for concreteness, we shall use a specific form which is widely used in the literature.
In general, the energy-density of a time-dependent classical axion field $a(t)$ is given by
\begin{equation} 
\rho_a =   {1 \over 2} \dot{a}^2 \, +  \, V(a/f_a) \,.        
\label{Edensity}
\end{equation} 
Since the purpose of this paper is to distinguish classical and quantum effects, we shall for a moment keep $\hbar$ explicit, while setting the speed of light equal to one.  Therefore, the fields and parameters  in (\ref{Edensity}) have the following dimensions: 
$[a] = [f_a] = \sqrt{(energy)/(time)}$ and $[V] = (energy)/(time)^3$.

Of course, for a generic non-singular function $V(a/f_a)$, the exact form of 
$a(t)$ can be very complicated and the oscillation-period can have a non-trivial  dependence on the amplitude. 
However,  as long as the axion field does not reach the maxima of the potential during its oscillations, the order of magnitude of the characteristic oscillation period $t_{\text{osc}}$ is given by the inverse curvature of potential 
$V(a)$ at its minimum, about which the axion oscillates. Without loss of generality, we can set the minimum to be at $a=0$.  
Then, $t_{\text{osc}}^{-2}  \sim {\partial^2 V(a) \over \partial^2a}|_{a=0}\, \equiv \, \bar{m}_a^2$, 
where $\bar{m}_a$ represents the frequency of small oscillations (with infinitesimal 
amplitude).  We put a bar on this quantity in order to distinguish it from its quantum counterpart $m_a$, which represents  the mass of the axion {\it particle} in the quantum theory.  The two quantities are related via $\hbar$ as $m_a \equiv  \hbar \bar{m}_a$.    

In the cosmological environment, the coherent oscillations of a homogeneous axion field are described by an equation similar to a damped anharmonic oscillator:
\begin{equation} 
\ddot{a} \, + \, \gamma \, \dot{a} \, + \, {\partial V(a) \over \partial a}  \,  = \, 0 \,,
\label{axionGeneric}
\end{equation} 
where the friction term $\gamma$ predominantly comes from the Hubble  
damping, $\gamma \simeq 3H$, due to the expansion of the Universe. The contribution from the axion decay is negligible. 
Note that for $\gamma \ll \bar{m}_a$, which is the case for most of the situations of our interest,   
we can still identify certain important properties of 
the time-evolution $a(t)$ without actually knowing the explicit forms of 
the functions $V(a)$ and $a(t)$.  The usual trick to achieve this is to first rewrite equation (\ref{axionGeneric}) in the 
following form:
\begin{equation} 
\dot{\rho_a} \, =  \, - \gamma \, \dot{a}^2 \,.
\label{average}
\end{equation} 
Next we can average this expression over a time-scale of order $\bar{m}_a^{-1}$, on which the variation of $\gamma$ is negligible and it can be treated as constant.  Moreover, if the axion oscillation amplitude is smaller than $f_a$, nonlinearities are not important and oscillations are dominated by the mass term. In such a case,  the average values over a period of oscillation of the kinetic and potential energies of the axion field  are equal and each carry half of the total energy density,   
${1 \over 2} \overline{\dot{a}^2} \, = \, \overline{V(a)} = {1 \over 2} \overline{\rho_a} $, so that we can replace $\overline{\dot{a}^2}$ on the r.h.s.\ of (\ref{average}) by  $\overline{\rho_a}$. 
Finally,  applying the resulting average expression for the evolution on time-scales longer than the Hubble-time $t \sim \gamma^{-1} \gg \bar{m}_a^{-1}$,  we get the following equation describing the time-evolution of the axion energy density: 
\begin{equation} 
\dot{\rho_a} \, =  \, - \gamma \, \rho_a  \,,
\label{avolved}
\end{equation} 
where we drop the bar from now on. This can be easily integrated to give 
\begin{equation} 
\rho_a(t)  \, =  \rho_a(t_{\text{in}})\, \exp\left(-\limitint_{t_{\text{in}}}^{t}  \gamma(t')\, \diff t'\right) \,, 
\label{densityEvolve}
\end{equation} 
where $t_{\text{in}}$ is some initial time. Taking into account 
that $\gamma = 3H$, we immediately get the well-known result that the axion energy density dilutes as the inverse-cube of the cosmological scale factor, i.e., 
redshifts just like dust.  

Correspondingly, we can use the temperature of the microwave background radiation in the Universe as a useful clock for keeping track of the axion energy density. Thus, we represent the evolution of the axion energy density in the following 
frequently-used form:
\begin{equation} 
\rho_a(T)  \, =  \rho_a(T_{\text{in}})\,  {T^3 \over T_{\text{in}}^3}  \,.  
\label{densityTemperature}
\end{equation} 
In the language of the oscillating classical field $a(t)$, this means that due to Hubble damping, the amplitude $A(t)$ of the axion field reduces in proportionality to $T^{3/2}$. The fact that the classical axion energy density redshifts as dust nicely matches the quantum intuition according to which the time-dependent classical axion field represents a mean field description of a quantum gas of cold bosons. 

In section \ref{sec:simplifications}, we will show that the friction term, which reduces the density of axions, does not affect the validity of the classical description: The axion evolution is still well-described by a classical solution of damped oscillations. 	
We shall therefore structure our analysis in the following way. First, we will
ignore the contribution from the friction term and develop a coherent-state 
picture of the axion in the absence of dilution.  In this setup, we identify the effects which lead to a quantum break-time of the axion field. We show that the time-scale is enormous.   
We shall later  take into account the underlying quantum effects which lead to friction in the classical theory and show that the original assumption that they do not contribute to quantum-breaking is consistent at the fundamental level. 

\subsubsection*{The Setup}
For concreteness,  we model the axion potential by the following widely-considered form: $V(a/f_a) = \Lambda^4 (1 - \cos(a/f_a)) $, where the scale $\Lambda$ set by QCD and quark masses has dimensionality $[\Lambda^4] = (energy)/(time)^3$. The time-dependent classical axion field then satisfies the following  equation:
\begin{equation} 
\ddot{a} \, + \, {\Lambda^4 \over f_a} \sin(a/f_a) \,  = \, 0 \,,
\label{axionEQ}
\end{equation} 
where we have set $\gamma =0$ as explained above. Now we assume that $a(t)$ is a solution  of (\ref{axionEQ}) which describes a would-be classical evolution of the system.  

Our task is to understand how strong the quantum effects are which 
give a departure from the classical description in terms of $a(t)$. Thus, we need to compare the full quantum evolution of the system with the classical one, i.e., we need to check how well the classical field $a(t)$ approximates the expectation value of the  quantized axion field 
$\hat{a}$ over its quantum state $\ket{N}$ which describes the true quantum evolution of the system.  To this end, we need to resolve the axion field in its quantum constituents. In order to make our bound maximally stringent, we shall choose as initial state of the axion field a coherent state, i.e., a quantum state which is maximally classical.  

The key difficulty in representing the axion field in form of a coherent state consists in identifying the proper creation and annihilation operators, 
$\hat{c}_{\vec{k}}^{\dagger} $, $\hat{c}_{\vec{k}}$. For the coherent state which describes anharmonic  oscillations, the creation and annihilation operators are the eigenstates of the interaction Hamiltonian and in particular do not satisfy the dispersion relation of free quanta. In order to circumvent this difficulty, we shall follow the treatment of \cite{us}. The idea is to first approximate the axion field by a free field, and then to construct the coherent state using the creation and annihilation operators of free quanta. For the quantum state constructed in this way, it is straightforward to calculate the quantum break-time $t_{\text{q}}$. 

However, the approximation as free field introduces a second time-scale, namely the classical break-time $t_{\text{cl}}$, after which classical nonlinearities become important. They modify the harmonic evolution but do not jeopardize the classicality of the description. Even though $t_{\text{cl}}$ is generically shorter than $t_{\text{q}}$, it is possible to extrapolate the quantum break-time to get an estimate for the corresponding time-scale in the full nonlinear theory. In order to show this, we make use of the general relation
	\begin{equation}
	t_{\text{q}} \, = \, {t_{\text{cl}} \over \alpha} \,,
	\label{relation}
	\end{equation}   
	where $\alpha$ is the quantum coupling. This formula was derived in \cite{us} and we shall explicitly reproduce it for the axion case. It reflects the fact that the two corrections are of fundamentally different nature and can be varied independently by appropriately adjusting $\alpha$. This means that also in a full nonlinear theory, the effects which lead to the quantum break-time occur on a similar time-scale as in the free theory. We refer the reader to \cite{us} for a more detailed discussion.

      \subsubsection*{Approximation as Free Field}
    
      In order to approximate the classical axion solution, we expand \eqref{axionEQ} in powers of $a/f_a$: 
       \begin{equation} 
      (\partial_t^2\, +  \, \bar{m}_a^2) a  -  { 1 \over 6} \bar{\alpha}_a a^3   \, + \, \ldots = 0 \,,      
      \label{eqaxion}
      \end{equation} 
       where $\bar{m}_a = \Lambda^2/f_a$ is the oscillation frequency and $\bar{\alpha}_a  \equiv {\bar{m}_a^2 \over f_a^2}$ is the nonlinear interaction strength. 
  For a small amplitude  $A \ll f_a$, the nonlinear terms can be ignored and we can approximate the solution by
 \begin{equation}
a_{0}(t) = A \cos (\bar{m}_a t) \,.
\label{eqn:freeSolution}
\end{equation}
 This is the solution which we will resolve as expectation value over an underlying quantum state. But before that, we determine the classical break-time, i.e., how long the solution \eqref{eqn:freeSolution} represents a valid approximation to the full nonlinear classical solution. 
We do so by estimating the first anharmonic correction $a_1$ to \eqref{eqn:freeSolution} in a series expansion in powers of $\bar{\alpha}_a$. Plugging in the split
\begin{equation}
	a = a_0 + \bar{\alpha}_a a_1
	\label{eqn:aSplit}
\end{equation}
in \eqref{eqaxion}, we obtain the equation of a driven harmonic oscillator:
\begin{equation}
(\partial_t^2 + \bar{m}_a^2 ) a_1 = \frac{A^3}{6} \cos^3(\bar{m}_a t) \,.
\label{eqn:drivenOscillator}
\end{equation}
Using the identity $\cos^3(x) = \left(\cos(3x) + 3 \cos(x)\right)/4$, it is easy to check that
\begin{equation}
	a_1 = A \frac{A^2}{192 \bar{m}_a^2} \left(\cos(\bar{m}_a t) - \cos(3 \bar{m}_a t) + 12\, \bar{m}_a t \,\sin(\bar{m}_a t)\right) 
	\label{eqn:A1}
\end{equation} 
is the solution for the initial conditions $a_1(0) = \partial_t a_1(0) = 0$. Thus, the leading deviation from the free solution is caused by the resonant term $\propto (\bar{m}_a t)$. It can be neglected as long as
\begin{equation}
	\bar{\alpha}_a \frac{A^2}{\bar{m}_a^2} \bar{m}_a t \ll 1 \,.
	\label{eqn:conditionClassicalBreakTime}
\end{equation} 
This leads to the classical break-time  
\begin{equation}
	t_\text{cl} =\bar{m}_a^{-1} \frac{\bar{m}_a^2}{\bar{\alpha}_a A^2}  \,,
	\label{eqn:classicalBreakTime}
\end{equation}
in accordance with equation (4) of \cite{us}. The nonlinear corrections which lead to this time-scale correspond to an expansion in the dimensionless quantity
\begin{equation}
\lambda \equiv 	\bar{\alpha}_a \frac{A^2}{\bar{m}_a^2} \,,
	\label{collective}
\end{equation} 
 which is classical in the sense that it is independent of $\hbar$.  It will become clear shortly that it corresponds to the collective coupling of axions in the quantum theory,  as defined in equation \eqref{collectiveGeneral}.  Already at this point, it is important to stress that corrections which scale like $\lambda$ correspond to classical nonlinearities and therefore cannot lead to quantum breaking.
 
 \section{Deviations from the Classical Evolution}
 \label{sec:quantum}
 
 \subsubsection*{A Quantum Description of the Classical Solution}

 We proceed to discuss the axion field on the quantum level. The fundamental Lagrangian, which corresponds to \eqref{eqaxion}, is 
  \begin{equation} 
    \hat{\mathcal{L}} =  {1 \over 2} \partial_{\mu} \hat{a} \partial^{\mu} \hat{a}\, -  \, { 1 \over 2\hbar^2} m_a^2 \hat{a}^2  +  {1\over \hbar 4!}  \alpha_a \hat{a}^4 \,.
  \label{axionL2}
  \end{equation} 
 Since we are in the quantum theory now, we switched to the relevant quantum quantities, such as the mass of the axion particle 
 $m_a \equiv \hbar \bar{m}_a$ and the dimensionless quantum coupling 
 $\alpha_a \equiv \hbar \bar{\alpha}_a$. 
 
 Our overall goal is to investigate corrections which lead to a departure of the true quantum evolution from the classical solution. To this end, the next step is to understand the classical solution as expectation value in an underlying quantum state. As explained, we will neglect interaction in doing so, $\alpha_a = 0$. 
   In that case, we expand the full Heisenberg operator $\hat{a}$ in creation and annihilation operators:
\begin{align}
\hat{a} = \sum_{\vec{k}} { \sqrt{\hbar  \over 2 V\omega_{\vec{k}}}} 
 \left(\hat{c}_{\vec{k}}e^{-ikx} + \hat{c}_{\vec{k}}^\dagger e^{ikx}\right) \,, \label{eqn:Exp:ModeExpansion}
\end{align} 
where $kx \equiv \hbar^{-1}\vec{k} \vec{x} - \omega_{\vec{k}} t$ and
$\omega_{\vec{k}} = \hbar^{-1}\sqrt{m_a^2 + |\vec{k}|^2}$.
The creation and annihilation operators satisfy the standard commutation relations $[\hat{c}_{\vec{k}},\hat{c}_{\vec{k}'}^\dagger] = \delta_{\vec{k},\vec{k}'}$. 

The simplest quantum state $\ket{N}$ of the scalar field $\hat{a}$ over which the expectation value is the classical harmonic oscillator solution $a_0(t)$ of equation \eqref{eqn:freeSolution} is a coherent state of zero momentum quanta:
\begin{align}
\ket{N} \equiv  \ \text{e}^{-\frac{1}{2}N + \sqrt{N} \hat{c}_{0}^\dagger }
\ket{0}\,  = &\, 
\text{e}^{-\frac{1}{2}N} \sum_{n=0}^\infty \frac{N^{\frac{n}{2}}}{\sqrt{n!}} \ket{n}  \,\text{, with}  \label{eqn:Exp:QuantumState}\\
N =&\ V {m_aA^2 \over 2 \hbar^2}  \,.\label{eqn:Exp:N}
\end{align}
In this formula, $\ket{n} = (\hat{c}_{\vec{0}}^\dagger)^n (n!)^{-\frac{1}{2}} \ket{0}$ are normalized number eigenstates of $n$ quanta with zero momentum. Because of (\ref{eqn:Exp:ModeExpansion}), it is obvious that the state $\ket{N}$ indeed yields the correct expectation value:
\begin{align}
\bra{N}\hat{a}\ket{N} =  A \cos(\bar{m}_at) \,,
 \label{eqn:Cla:ExpectationValue}
\end{align}
where we used that $\hat{c}_{\vec{0}} \ket{N} = \sqrt{N} \ket{N}$. From the point of view of the non-interacting quantum theory (i.e., for $\alpha_a = 0$), the oscillating classical axion field $a_0(t)$ therefore is a coherent state of zero-momentum axions of mean occupation number density
\begin{equation}
   n_a = m_aA^2\hbar^{-2} \,, 
   \label{occupationDensity}
\end{equation}
 where we dropped an irrelevant factor of $2$. Consequently, the energy density is
\begin{equation}
	\rho_a = m_a^2A^2\hbar^{-2} \,.
	\label{energyDensity}
\end{equation}   
If we restrict ourselves to the volume $V = \bar{m}_a^{-3}$, we get the mean occupation number 
    \begin{equation}
    N= \hbar^{-1} {A^2 \over \bar{m}_a^2} \, .   
    \label{N}
    \end{equation}
     As already discussed in section \ref{sec:scaling}, we observe that $N$ scales as $\hbar^{-1}$. This means that it becomes infinite in the classical limit $\hbar \rightarrow 0$.
    Furthermore, we see that the quantity (\ref{collective}) indeed has the meaning of the collective coupling in the quantum language: $\lambda = \alpha_aN$. Since we have switched to the quantum theory,  we can set 
    $\hbar =1$ from now on. 
  
   \subsubsection*{Inclusion of Quantum Coupling}  
    As soon as we consider $\alpha_a \neq 0$, the interaction terms create quantum Hamiltonian processes which lead to a departure from the coherent state on the time-scale of  the quantum break-time $t_{\text{q}}$.
   After this time, the initial coherent state generically will evolve in something non-coherent and the expectation value of the quantum field will no longer match the classical solution. Since we want to put a lower bound on $t_{\text{q}}$, it is not important for us what the resulting ("departed") state will be.  We  just want to figure out the shortest time $t_{\text{q}}$ before the coherent state approximation and therefore the classical description can break down.

     This can be estimated  in the following way. 
     The Hamiltonian processes which lead to a departure from the coherent state are due to the rescattering of axions. The simplest such process would be the scattering due to four-point coupling. However, 
     in the approximation in which the initial and final particles are treated as free, the rate is suppressed due to momentum conservation: Since both initial and final axions carry vanishing momentum, the phase space for such a process is zero.  
 
\begin{figure}
   	\begin{center}
   		\includegraphics[width=0.7\textwidth]{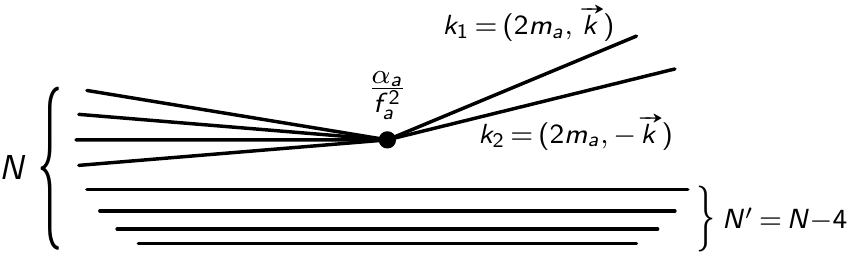}
   		\caption{Four coherent state axions annihilate into two axions of 4-momentum $k_1$ and $k_2$, whose momenta are non-zero. }
   		\label{fig:Scattering42}
   	\end{center}
   \end{figure}

   Therefore, the lowest order non-vanishing process involves the participation of six axions.  For instance, four  coherent state axions can annihilate into two axions with non-zero momenta, as is depicted in figure \ref{fig:Scattering42}. 
    The final state of such a process can be a tensor product of two one-particles states of 4-momenta $k_1=(E_1,\vec{k}_1)$ and $k_2=(E_2,\vec{k}_2)$ with a coherent state $\ket{N-4}$ of zero momentum axions of reduced mean occupation number $N-4$. 
By energy-momentum conservation, we have $\vec{k_1} = -\vec{k_2}$ 
and $E_1 = E_2 = 2 m_a^2$.  The tree-level matrix element for the transition  $\ket{N} \rightarrow \ket{N-4}\ket{\vec{k_1}} \ket{\vec{k_2}}$ comes  from six-point interaction in the Hamiltonian (or from four-point interactions with one virtual axion exchange):
  \begin{equation} 
 {\alpha_a\over f_a^2}\bra{N-4} \bra{\vec{k_1}} \bra{\vec{k_2}} 
( \hat{c}_{\vec{k_1}}^{\dagger}\hat{c}_{\vec{k_2}}^{\dagger} \hat{c}_0\hat{c}_0  
\hat{c}_0\hat{c}_0 )\ket{N}   = {\alpha_a\over f_a^2}N^2\ex^{-{2 \over N}} \,, 
    \label{process} 
   \end{equation}  
    where the factor $\ex^{-{2 \over N}}$ is due to the overlap of coherent states with different mean occupation number. For large $N$, we can approximate it as $\ex^{-{2 \over N}} \approx 1 - 2/N$. Thus, a different final coherent state leads to a correction which scales like $1/N$. The rate of the scattering process is: 
         \begin{equation} 
\Gamma_{4\rightarrow 2}  \sim  m_a (\alpha_a N)^4  =  m_a \lambda^4\,, 
\label{rate}  
\end{equation}
 where we used the collective coupling $\lambda$ as defined in \eqref{collectiveGeneral}. 
 This is the rate at which the axion coherent state loses its constituents 
 and \textit{de\hyp{}classicalizes}.  Roughly speaking, for a small enough $\Delta N$, 
 the rate of a process in which the coherent state looses $\Delta N$ constituents goes as $\Gamma_{N \rightarrow (N - \Delta N)}  \sim   m_a \lambda^{\Delta N}$.

  Note that for tree-level multi-particle amplitudes with a large enough number of axion legs, perturbation theory is expected to  break down due to a factorial growth of the diagrams.  
   This breakdown will not affect our conclusions because of the following two reasons.  First, non-perturbative arguments indicate that multi-particle quantum processes -- e.g., a process in which the axion coherent state looses order-one fraction of its constituents during one oscillation time -- must be exponentially suppressed.   
   
   Secondly, the breakdown of perturbation theory due to the  factorial growth  of diagrams only takes place for a number of axion 
 legs exceeding $\alpha_{a}^{-1}$, i.e., they occur for $N > \alpha_{a}^{-1}$ or equivalently  $\lambda >1$. 
  Such amplitudes  are outside of  the domain of our interest since -- as it will become clear  below --
the occupation number of axions,  which could  serve as a viable dark matter candidate,  is much smaller than this perturbative bound. 
In other words, for axions we have $N \ll  \alpha_{a}^{-1} $ and $\lambda \ll 1$. 
In this domain, multi-particle processes, in which e.g.,
of order $N$ cold axions annihilate into few  energetic ones, 
can be reliably computed and are 
exponentially suppressed.
The exponential suppression of such processes 
 can be explicitly seen by adopting the results of  \cite{graviton}, 
 where analogous multi-particle amplitudes have been computed for 
 gravitons.     For reading out the exponential suppression, the 
 difference between axions and gravitons is inessential.

Hence, we can safely conclude that the leading order process which leads
to de\hyp{}classicalization of the spatially-homogeneous time-dependent axion field  has the rate (\ref{rate}).

\subsubsection*{Quantitative Estimates}

 In order to make quantitative estimates, it is useful to express the rate in terms of the oscillation amplitude and the decay constant: 
       \begin{equation} 
\Gamma \,   = \, m_a  {A^8 \over f_a^8} \,.
\label{rate1}  
\end{equation}
Since this is the rate of the leading process, we have dropped the subscript. 
We recall that according to (\ref{densityTemperature}), the axion energy density drops as temperature-cubed because of Hubble friction. Therefore, it follows from \eqref{energyDensity} that the amplitude decreases as $A \propto T^{3/2}$.  Expressing $A$ in (\ref{rate1}) as function of temperature and evaluating it for today's value $T = T_{\text{today}}$,  
we get a minuscule rate for the decay of the axion coherent state in the present epoch:
\begin{equation} 
\Gamma \sim m_a \left ({T_{\text{today}} \over T_{\text{in}} } \right )^{12} \sim m_a\, 10^{-144}\,,
\label{gamma} 
\end{equation}
 where $T_{\text{in}} \sim 100\,$MeV is taken as QCD phase transition temperature and we also conservatively assumed $A_{\text{in}} \sim f_a$.   
  Thus, the characteristic rescattering time required to reduce the coherence of today's axion field by a factor of order $1/N$ already exceeds the age of the Universe by many orders of magnitude.  

However, the time-scale $\Gamma^{-1}$ is not yet the quantum break-time. The quantum break-time $t_{\text{q}}$ is much longer. Since the mean occupation number in the coherent state is $N$, quantum rescattering is able to  give a significant departure from the coherent state only after a sufficiently large fraction of particles (i.e., of order $N$) experience rescattering. 
This process takes the time
       \begin{equation} 
t_{\text{q}} = N\,\Gamma^{-1}  = m_a^{-1}  {f_a^8 \over m_a^2 A^6} \, .
\label{time}  
\end{equation}
We see that the quantum break-time increases as $A^{-6}$. Let us evaluate it at different epochs.  
First, we consider it at the onset of oscillations assuming the initial amplitude to 
 be maximal: $A_{\text{in}}=f_a$. 
For example, taking the axion mass $m_a \sim 10^{-5}\,$eV, which implies $f_a \sim 10^{12}\,$GeV, we get $t_{\text{q}} \sim 10^{52}\,$cm. Even in this crude estimate, in which we ignore Hubble dilution, the quantum break-time exceeds the current age of the Universe by a factor of approximately $10^{24}$.  
	
 In order to estimate the quantum break-time \eqref{time} for the present epoch, we can express it  in the form
	\begin{equation}
		t_\text{{q}} = m_a^{-1}  \frac{f_a^8 m_a^4}{\rho_a^3} \,, 
		\label{time4to2}
	\end{equation}
	where we plugged in the energy density \eqref{energyDensity}. Using the energy density of dark matter, $\rho_a \sim (10^{-3}\,\text{eV})^4$, we obtain
	$t_{\text{q}} \sim 10^{184}\,$cm,	
	which exceeds the current age of the Universe by a factor of approximately
	$10^{156}$. 
	Thus, the coherent  state approximation for describing the axion field in the present epoch is extremely accurate.
	
     \subsubsection*{Effect of Non-Zero Axion Momenta}
   So far, we have only considered $2 \rightarrow 4$-scattering. In reality, since the classical axion field in the Universe is a distribution over different wavelengths,  the $2 \rightarrow 2$-rescatterings can also contribute to decoherence.
  For completeness, we therefore give an estimate of the quantum break-time due to such processes.  
 A coherent state of axions which includes different momenta has the form
 \begin{equation} 
\ket{a} =  \prod_{\vec{k}}\otimes \ket{N_{\vec{k}}} \,, \label{difK}
 \end{equation}
  where a direct product is taken over the coherent states 
  $\ket{N_{\vec{k}}} $ of different momenta 
  $\vec{k}$ and mean occupation numbers  $N_{\vec{k}}$:    
 \begin{equation}
\ket{N_{\vec{k}}} \equiv  \ \text{e}^{-\frac{1}{2}N_{\vec{k}} + \sqrt{N_{\vec{k}}} \hat{c}_{\vec{k}}^\dagger }
\ket{0}\,.
\label{CohK}
\end{equation}
 Of course, since axion dark matter is cold, most of the energy is carried by axions with $|\vec{k}| \ll m_a$.  However,  in order to make our bound more conservative, we shall allow the axion momenta to be comparable to  $m_a$, which would make the rescattering rate higher.   
 In that case, the two body rescattering rate is given by
      \begin{equation} 
\Gamma_{2\rightarrow 2} \,   = \, m_a \alpha_a^2 N^2 = m_a  {A^4 \over f_a^4}  \,.
\label{rate2}  
\end{equation}
For today's axion density, we get 
$\Gamma_{2\rightarrow 2} \sim m_a\, T_{\text{today}}^6 / T_{\text{in}}^6 \sim m_a\, 10^{-72}$,  which is again minuscule.

The corresponding quantum break-time is given by
       \begin{equation} 
t_{\text{q}} = N\, \Gamma_{2\rightarrow 2}^{-1}  = m_a^{-1}  {f_a^4 \over m_a^2 A^2} = 
m_a^{-1} {f_a^{4} \over \rho_{a}} \, .
\label{time2to2}  
\end{equation} 
 Evaluating this expression at the onset of oscillations with maximal amplitude 
 $A = f_a$, we get the same result as for the $4\rightarrow2$-scattering: $t_{\text{q}} \sim 10^{52}\,$cm. On the other hand, if we evaluate the quantum break-time for the current epoch, in which the axion energy density is taken to be the dark matter density, we get $t_{\text{q}} \sim 10^{96}\,$cm,
 which exceeds the current age of the Universe by a factor of approximately $10^{68}$.  Thus, the inclusion of $2\rightarrow2$-scattering by taking into account the distribution of coherent state axions over different momenta 
  does not change our conclusion that the coherent state description for dark matter axions is extremely accurate.  
 
  Finally, we note that comparing the result \eqref{time2to2} to the classical break-time \eqref{eqn:classicalBreakTime}, we recover relation \eqref{relation}, $t_{\text{q}} = t_{\text{cl}}/\alpha$, for the process of $2\rightarrow2$-scattering.

 \section{Validity of Our Simplifications}
 \label{sec:simplifications}
 
  In order to conclude our argument, we wish to point out that the simplifying assumptions which we have made in our estimates do not significantly change the exact result. For example, one could wonder whether particle production, which takes place in an expanding universe, could have any significant effect. In particular, one might be worried about produced free axions, which could scatter off the coherent axion state and lead to decoherence. However, this does not happen since not enough particles are produced: Because of $\gamma = 3 H \ll m_a$, particle production is exponentially suppressed by the Boltzmann factor $\exp(-m_a/H)$.
 
 A similar argument holds for the effect of the QCD phase transition, which leads to a change of the axion mass. The corresponding transition time $t_{\text{QCD}}$, which is of order of the Hubble 
 time around the QCD-temperature, $t_{\text{QCD}} \sim  {M_P \over 
 	\Lambda^2}$, is much longer than the axion Compton wavelength: $t_{\text{QCD}} \gg m_a^{-1}$. Therefore, $\dot{m}_a \ll m^2_a$, i.e., the transition is adiabatic. Consequently, the quantum creation of free axions due to the time dependence of the mass is suppressed by $\exp(-m_a^2/ \dot{m}_a)$ and there are not enough produced particles to have a significant effect on the coherence of the axion state.
 
 Finally, we turn to the dilution of the axion number-density. In a  cosmological context, this dilution originates from the Hubble expansion as well as the decay of axions into some lighter particle species, e.g., photons. 
 For realistic values of the axion mass, the decay is a subdominant process. Therefore, we focus on the effect of Hubble friction. Since it describes a classical process, namely the dilution of gas in the background of an expanding universe, we intuitively expect that it does not lead to quantum decoherence. 
 
 Our goal, however, is to make this statement more precise. Thus, we generalize Lagrangian \eqref{axionL2}:
\begin{equation}
\hat{\mathcal{L}} = \frac{\textswab{a}^3}{2}  \left(\partial_\mu \hat{a} \partial^\mu \hat{a}  - m_a^2 \hat{a}^2\right) \,,
\end{equation}
where $\textswab{a}$ is the scale factor. We do not consider self-interaction of the axion since our present goal is only to investigate possible decoherence due to Hubble friction. The canonically conjugate momentum $\hat{\Pi}_a = \textswab{a}^3 \partial_0 \hat{a}$ gives the Hamiltonian
\begin{equation}
\hat{\mathcal{H}} = \frac{\hat{\Pi}_a^2}{2 \textswab{a}^3} +  \frac{\textswab{a}^3}{2} \left((\textswab{a}\vec{\partial} \hat{a})^2 + m_a^2 \hat{a}^2\right)\,.
\end{equation}
Specializing to a spatially homogeneous field, the Heisenberg equation of motion is
\begin{equation}
\ddot{\hat{a}} + \gamma \dot{\hat{a}} + m^2\hat{a} =0 \,,
\label{dampedHeisenbergEom}
\end{equation}
since $\gamma = 3H = 3\dot{\textswab{a}}/\textswab{a}$. As the Heisenberg equation is linear, we can apply the Ehrenfest-theorem, i.e, we can take its expectation value to conclude that for any quantum state, the expectation value of the time-evolved quantum operator is equal to the classical solution given by \eqref{axionGeneric}. Thus, the fact that friction is a linear term suffices to show that it cannot destroy classicality.

For completeness, we will nevertheless explicitly study the quantum time evolution, where we use the calculation of \cite{chenFung}.
In the Schr\"odinger picture, this paper derives the following eigenvalue equation for an initially coherent state $\ket{N}$:
\begin{equation}
\left(\text{e}^{imt}\cosh(\frac{\gamma t}{2})\, \hat{c}_{\vec{0}} + \text{e}^{imt}\sinh(\frac{\gamma t}{2})\, \hat{c}_{\vec{0}}^\dagger \right)\ket{N(t)} = \sqrt{N} \ket{N(t)} \,,
\label{evolvedEigenvalue}
\end{equation}
where the creation and annihilation operators $\hat{c}_{\vec{0}}^\dagger$, $\hat{c}_{\vec{0}}$ are still defined by the mode expansion \eqref{eqn:Exp:ModeExpansion} (at $t=0$).
In order to make the formulas more transparent, we wrote down the solution only to leading order in $\gamma/m$ and neglected $\dot{\gamma}/(m\gamma)$. As explained above, both simplifications are reasonable in a cosmological context. In this limit, \eqref{evolvedEigenvalue} shows that $\ket{N(t)}$ is a squeezed coherent state with real squeezing parameter $s= \gamma t/2$. The uncertainty is no longer equally distributed:
\begin{equation}
\Delta a \propto \text{e}^{-\gamma t/2} \,, \ \ \ \Delta \Pi \propto \text{e}^{\gamma t/2} \,,
\end{equation}
but still minimal.\footnote
{The emerging physical picture has a straightforward interpretation. In terms of the physical field $\partial_0 \hat{a}$, we have
	\begin{equation}
	\Delta a = \Delta \partial_0 a \propto  \text{e}^{-\gamma t/2} \,.
	\end{equation}  
	From this point of view, $\ket{N(t)}$ is therefore still coherent. Time evolution only reduces the overall uncertainty in physical space.
		This is in accordance with the commutation relations $\left[\hat{a}(t,\vec{x}), \hat{\Pi}_a(t,\vec{y})\right] = i \delta(\vec{x}-\vec{y})$, which read in physical space
		\begin{equation}
		\left[\hat{a}(t,\vec{x}), \partial_0 \hat{a}(t,\vec{y})\right] = i  \textswab{a}^{-3} \delta(\vec{x}-\vec{y}) \,.
		\label{commutationRelationsPhysical}
		\end{equation}
	This means that uncertainty is conformally conserved but dilutes with the physical volume.}
Thus, $\ket{N(t)}$ continues to be maximally classical.
Including higher terms in $\gamma/m$ does not change these conclusions significantly.\footnote
{In this case, the state $\ket{N(t)}$ is no longer exactly squeezed at all times, but the deviation of $\Delta a \Delta \Pi$ from 1 vanishes periodically and is bounded by a constant which scales as $\gamma^2/m^2$.}
This means that also in the presence of classical Hubble friction, the classical description of the free axion field remains valid indefinitely. It does not lead to a quantum break-time.

\section{Subcriticality of Axion Gas} 
\label{sec:collective}

\subsubsection*{The Role of Classical Instabilities in Fast Quantum Breaking}
 As we have already discussed in section \ref{sec:scaling}, it was demonstrated in \cite{nico} that a simple system of cold bosons can exhibit a remarkably short quantum break-time in the regime in which the boson gas is overcritical and unstable. In that case, this time-scale is determined by the Lyapunov exponent of the classical instability, which the system exhibits in this regime.
 
  In this section, we would like to clarify what such a regime would imply for the axion field and how far the cosmic dark matter axions are from criticality. As a useful reference point,  we shall confront the axion gas 
  with the quantum gas studied in \cite{nico}. This system consists of non-relativistic bosons of mass $M$, which are contained in a periodic $d$-dimensional box of radii $R$ and exhibit a simple attractive interaction. The Hamiltonian has the following form:
   \begin{equation}
 \mathcal{\hat{H}} \, = \, \int \ldiff[d]{\vec{x}} \, \hat{\psi}^{\dagger} \frac{- \hbar^2 \Delta}{2M} \hat{\psi} \, - \, \frac{g}{2}   \ 
  \int \ldiff[d]{\vec{x}} \, \hat{\psi}^{\dagger}\hat{\psi}^{\dagger} \,\hat{\psi} \hat{\psi} \,, 
\label{Hnonderivative} 
\end{equation} 
where the parameter $g > 0$ controls the strength of the {\it attractive} interaction among the bosons.  
 
 We can represent $\hat{\psi}$ as a classical mean field and the quantum fluctuations, 
 $\hat{\psi} \, = \, \psi_{\text{cl}} +  \hat{\delta \psi}_{\text{q}}$, where $\psi_{\text{cl}}$ satisfies the Gross-Pitaevskii equation: 
 \begin{equation}
  i \hbar \partial_t \psi_{\text{cl}}    = 
 \left( -\frac{\hbar^2}{ 2M} \Delta    -   g |\psi_{\text{cl}}|^2
  \right  ) \psi_{\text{cl}} =   
  \mu \psi_{\text{cl}} \,.  
   \label{GPeq}
 \end{equation}
The parameter $\mu$ is the chemical potential, which plays the role of a Lagrange multiplier for imposing the constraint $\int \ldiff[d]{\vec{x}} |\psi_{\text{cl}}|^2 =  N$. We shall focus on the homogeneous solution, 
   \begin{equation}
   |\psi_{\text{cl}}|^2 =  -{\mu \over g} = {N \over V} \,,   
   \label{homogeneousSolution}
 \end{equation}
 where $V\, = \, (2\pi R)^d$ is the $d$-dimensional volume. It represents a mean field description of the quantum state in which only the zero-momentum mode is macroscopically occupied. This solution exists for all non-zero values of the parameters. However, beyond a certain critical point it becomes unstable and the system undergoes a quantum phase transition.  In the overcritical regime, the instability of the homogeneous solution manifests itself as Lyapunov exponent which is independent of $\hbar$.
 
  The  role of the criticality parameter is played  by the effective collective coupling 
 \begin{equation}  
 \lambda_{\text{nr}}  \equiv  4gN \frac{M R^2}{\hbar^2 V}\,,
 \label{colNR}
 \end{equation} 
 where the subscript $nr$ serves to distinguish it from the analogous collective coupling in the axion case.   In order to investigate stability, we go to momentum space: $\hat{\psi} \, = \, \sum_{\vec{l}} \frac{1} {\sqrt{V}} {\rm e}^{i  {\vec{l} \over R} \vec{x}} \, \hat{a}_l$. Here $\vec{l}$ is the $d$-dimensional wave-number vector, which determines the momentum as $\vec{k} = \hbar {\vec{l} \over R}$.  The operators $\hat{a}_{\vec{l}}^\dagger, \hat{a}_{\vec{l}}$ are the creation and annihilation operators of bosons of momentum-number vector
${\vec{l}}$ and satisfy the usual algebra: $[\hat{a}_{\vec{l}}, \hat{a}_{\vec{l'}}^\dagger] = \delta_{\vec{l}\vec{l'}}$
and all other commutators zero. Then the Bogoliubov-de Gennes frequencies   are given by \cite{bogoliubovDeGennes}:
\begin{equation}
\hbar \omega_{\vec{l}} =  |\vec{l}  
|\frac{\hbar^2}{ 2M R^2} \sqrt{(|\vec{l}|^2 - \lambda_{nr})} \,.
\label{omega}
\end{equation} 
They become imaginary for all modes with $|\vec{l}|^2 < \lambda_{\text{nr}}$.  Thus, instability sets in for $\lambda_{\text{nr}}>1$, 
and the number of unstable $l$-modes depends on the magnitude of 
the criticality parameter $\lambda_{\text{nr}}$. 
In the regime in which only the $|\vec{l}|=1$-mode is unstable, the explicit analysis of \cite{nico} shows that the quantum break-time scales as $t_{\text{q}} \sim \omega_{\vec{1}}^{-1} \ln(N)$. 
By increasing $\lambda_{\text{nr}}$, one can destabilize higher and higher momentum modes and correspondingly make the Lyapunov exponents large.   Finally we note that $\omega_{\vec{l}}$ are classical quantities because $\hbar/M$ is classical and represents a zero mode oscillation frequency of an underlying classical field, whose quanta are the bosons in question. Thus, an important message which we take from the 
 results of \cite{nico} recounted above is: The quantum break-time can be shortened in the overcritical regime, provided that the initial state 
exhibits a classical instability, i.e., an instability characterized by an
$\hbar$-independent Lyapunov exponent.  

However, the following  clarification is in order. The expression (\ref{omega}) creates the impression that we can make the quantum break-time arbitrarily short time if the system is sufficiently overcritical, i.e., if we increase the collective coupling $\lambda_{\text{nr}}$, e.g., by keeping all the other parameters fixed and increasing the occupation number of zero-momentum quanta. However, one has to be very careful with this limiting case. Although in the non-relativistic model given by the Hamiltonian (\ref{Hnonderivative}) taking the limit $\lambda_{\text{nr}} \rightarrow \infty$ is legitimate, an underlying fundamental relativistic quantum field theory may go out of the validity.  For example, in the case of cold bosons, one reason is that they can only be described by the Hamiltonian (\ref{Hnonderivative}) as long as the gas is  sufficiently dilute.
  As we shall discuss below, similar limitations prevent the axion field from entering into a deep overcritical regime. 
    
  \subsubsection*{Level of Criticality of Cosmic Axions}
   In order to identify the instability domain of  the axion model, we first consider the truncated theory (\ref{axionL2}).   
 The potential has a minimum at $a=0$, a maximum at 
 $a_{\text{cr}}^2 = 6 {m_a^2 \over \alpha_a} = 6 f_a^2$ and is unbounded from below for larger values of $a$.  Also here, the collective coupling (\ref{collective}) of zero momentum axions  plays the role of the criticality parameter (with critical value $\lambda = 6$).  Overcriticality of the axion gas would mean that the amplitude of oscillations $A$ exceeds $a_{\text{cr}}$ and the classical field 
 $a_{\text{cl}}(t)$ grows unbounded.  This classical growth is accompanied by an instability of  modes with momenta  $ |\vec{k}| < \sqrt{{\alpha_a \over 2}  a_{\text{cl}}(t)^2 - m_a^2}$ so that finally all the modes become unstable. Of course, in such a situation the quantum breaking can be efficient, but it is also meaningless since in this regime 
 the truncated model no longer describes physics of the axion gas correctly.  The truncation is only meaningful as long as the amplitude of oscillations does not exceed $a_{\text{cr}}$. So the short quantum breaking in a would-be overcritical regime of axion gas is unphysical 
and is an artifact of an invalid description.  
    
   Let us now go to the full axion model with periodic potential, where we use as before $V(a) = \Lambda^4 (1- \cos(a/f_a))$.\footnote
   {Naturally, in doing so, we ignore a possible back reaction 
   	from the axion field on QCD dynamics.} 
   We can make the axion overcritical by assuming a high number density of zero momentum axions. 
   In the classical language, this means that we are looking for a time-dependent solution with energy density $\rho_a \gg \Lambda^4$.  In this regime, the solution, up to corrections ${\mathcal O} (\Lambda^4/\rho_a)$, has the form: 
  \begin{equation} 
   a_{\text{cl}}(t) =   \sqrt{2\rho_a} t \, .  
  \label{freeT} 
\end{equation}     
This solution has an obvious physical meaning. Since the energy of the axion field exceeds the height of the axion potential, the evolution is the one of a free field with a constant energy density. Let us examine the stability of this solution. The momentum modes of the linearized perturbations around it satisfy the following equation: 
  \begin{equation} 
   \ddot{a}_{\vec{k}}(t)  +  \left (\hbar^{-2}\vec{k}^2  + \bar{m}_a^2 {\rm cos}\left(  \sqrt{2\lambda} \bar{m}_at \right ) \right) a_{\vec{k}}(t) = 0 \,,
  \label{Meq} 
\end{equation}      
   where we have used the  collective coupling in the form
   \begin{equation}
   \lambda = {\rho_a \over f_a^2 m_a^2} \,,
   \label{collectiveRho}
   \end{equation}
  which follows from \eqref{collective} when we express the amplitude $A$ in terms of the energy density \eqref{energyDensity}. Introducing a new variable  $y \equiv \sqrt{{\lambda \over 2}}\bar{m}_a t$, 
 we can rewrite (\ref{Meq}) in the form
   \begin{equation} 
   \partial_y^2 a_{\vec{k}}(y)  +  \left (A  + {2 \over \lambda}  \cos(2y) \right) a_{\vec{k}}(y) = 0 \,,  
  \label{Meq1} 
\end{equation}   
where $A \equiv  {2\vec{k}^2 \over \lambda m_a^2}$. 
 
This is the Mathieu equation, which is known to exhibit instability bands around certain values of $A$ \cite{mclachlan_theory_1951}. However, it is important to remember that we work in the approximation $\rho_a \gg \Lambda^4$, which is equivalent to a large collective coupling: $\lambda \gg 1$. Therefore, the term $\propto \cos(2y)$ responsible for generating the instability is suppressed. For this reason, the instability bands get narrower as $\lambda$ increases so that the phase space for the production of the corresponding modes is suppressed.\footnote
{In this regime, the most relevant instability is the lowest-lying one around $A=1$, i.e., for modes with $\vec{k}^2 \approx \lambda m_a^2/2$.}

\begin{figure}
 	\begin{center}
        \includegraphics[width=0.6\textwidth]{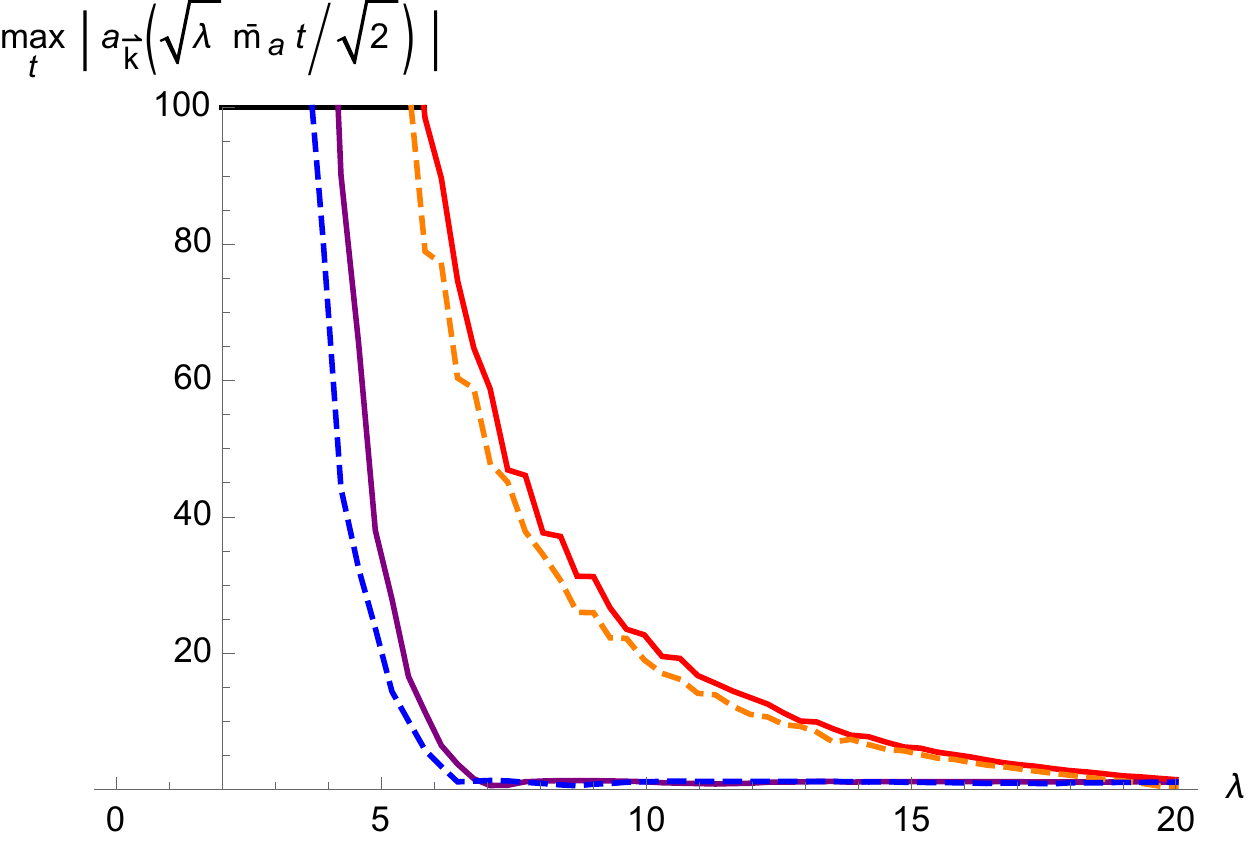}
 		\caption{ Behavior of the solution of \eqref{Meq1} for fixed $t$ and $A$ when the collective coupling $\lambda$ changes. The maximal value of $ a_{\protect \vec{k}}|(\sqrt{\lambda/2}\bar{m}_a t)|$ in the interval $20 \leq\bar{m}_a t \leq 25$ is plotted. The values for A are 0.85 (purple), 0.95 (red), 1.05 (orange, dashed) and 1.15 (blue, dashed). The closer $A$ is to the critical value $1$, the stronger the instability is and the longer it persists. In each case, the instability disappears for big $\lambda$.}
 		\label{fig:mathieu}
 	\end{center}
 \end{figure}

Additionally, we study the scaling of the time-scale of instability. To this end, we investigate how an unstable solution $a_{\vec{k}}(\sqrt{\lambda/2}\bar{m}_a t)$ of \eqref{Meq1} changes at a fixed $t$ when we increase $\lambda$. In doing so, we change $\vec{k}$ in order to keep $A$ fixed near an unstable value. Numerical analysis shows that the time-scale of instability for a given mode indeed becomes longer as the collective coupling $\lambda$ increases, i.e., the instability disappears for $\lambda \rightarrow \infty$. This is not surprising since \eqref{Meq1} becomes the equation of a free particle in this limit. The result for the dominant instability around $A=1$ is displayed in figure \ref{fig:mathieu}. Therefore, we conclude that the increase of stability caused by the narrowing of the instability bands outweighs the reduction of the break-time due to the scaling $t\propto y/\sqrt{\lambda}$. In summary, this shows that once the full axion potential is taken into account, the quantum break-time cannot be made arbitrarily short even in the overcritical regime.

 As a final remark, we note that $\dot{a}$ cannot be arbitrarily large due to the fact that it back-reacts on the Peccei-Quinn field.  In particular, there is an absolute bound on $\dot{a}$ given by 
  $\dot{a} \sim f_a^2$ because at this point, the back reaction from the axion field 
  on the VEV of the modulus of the Peccei-Quinn field,  $\Phi_{PQ} \, \equiv  
  f_a \text{e}^{i {a \over f_a}}$, becomes order one and the axion decay constant $f_a$ changes. This must be taken into account.  So the simple description in terms of a pseudo-scalar $a$ with a periodic potential breaks-down and one has to consider the full theory.  For the collective coupling of the axion gas, this restriction  translates as the bound $\lambda < \alpha^{-1}_a$. 
  
  In conclusion, in the overcritical domain the quantum break-time
 can in principle be made shorter at the expense of Lyapunov instabilities
 along the lines of  mechanism of \cite{nico}.   However, 
 this domain is irrelevant for the cosmic axion field because of the following reasons.  
 First, this regime cannot be reached within the validity of axion effective field theory model (\ref{axionL2}), due to back reaction. 
  Secondly, the would-be overcritical domain -- in which potentially a fast quantum breaking could occur -- is way outside of the realistic parameter 
  space of dark matter axions in our Universe.   
The axion gas is clearly under-critical. For example, using the present energy density of dark matter as for equation \eqref{time4to2}, we get from \eqref{collectiveRho}: $\lambda = 10^{-44}$, which is minuscule.

\section{Relationship to Other Work}
\label{sec:otherWork}

We conclude by discussing in more detail the relationship of our work to the results presented in \cite{sikivieFirst, sikivieDetailed, sikivieAngular}. Also there, the classical axion field is resolved as a multi-particle quantum state. Subsequently, the authors investigate the axionic self-interaction. They do so by calculating the process of $2\rightarrow2$-scattering, which we also considered. In full agreement with our result, they obtain the quantum break-time \eqref{time2to2} (see $\Gamma_s$, which is defined before equation (8) of \cite{sikivieFirst}). Therefore, they also conclude that the time-scale of this process vastly exceeds the age of our Universe and does not play any role for current observations.
 They proceed, however, to study processes of $2\rightarrow2$-scattering in which also the final state is macroscopically occupied. Clearly, this enhances the scattering rate by $N$ and leads to the time-scale
\begin{equation}
	\tilde{t} = \frac{t_q}{N} = \bar{m}_a^{-1}\frac{1}{(\alpha_a N)^2}\,.
	\label{timeSikivie}
\end{equation}
At the onset of oscillations, it can indeed be short: $\tilde{t} = \bar{m}_a^{-1}$. They argue that this process leads to Bose-Einstein condensation of the axions, i.e., they increasingly occupy the mode of zero momentum.\footnote
{Our approach is even more radical  with respect to the distribution in momentum space. We already start with a fully condensed state in which all axions are in the mode of zero momentum.}

We are, however, not interested in implications of this process. Our key point is that $\tilde{t}$ only depends on the collective coupling $\lambda=\alpha_a N$ (see equation \eqref{collectiveGeneral}) and therefore is a \textit{classical} quantity independent of $\hbar$. It is the time-scale of classical nonlinearities, i.e., it has the same status as the classical break-time \eqref{eqn:classicalBreakTime}. Thus, the existence of this time-scale  does not jeopardize the classical description of the axion field or lead to a quantum break-time. This agrees with the discussion in reference \cite{condensation}, which is cited in \cite{sikivieFirst}.  Also there, it is noted that the process of condensation corresponds to a classical interaction of different momentum modes and can be described as scattering of classical waves.\footnote
{ That condensation can be described classically was also discussed more recently in \cite{guthCondensation}.}

As a second step, \cite{sikivieFirst, sikivieDetailed, sikivieAngular} contains the study of gravitational self-interaction of the axions. It is argued that it becomes strong at late times. Also there, short time-scales only appear when classical processes are considered, i.e., ones in which also the final state of scattering is macroscopically occupied.\footnote
{We can explicitly conclude this from equation (11) of \cite{sikivieFirst}. Writing  the time-scale of gravitational interaction as
    \begin{equation}
    \tilde{t}_g = \bar{m}_a^{-1} \frac{1}{\alpha_g N} \,,
    \end{equation}
    where $\alpha_g = \hbar G m_a^2$, and taking into account that $N$ scales like $\hbar^{-1}$, we conclude that $\tilde{t}_g$ is independent of $\hbar$, i.e., classical.}
As before, we do not want to make any statement about these effects.\footnote
{We would be surprised, however, if the gravitational self-interaction were strong. If we look at e.g., the cross section in equation (3.30) of \cite{sikivieDetailed},
   \begin{equation}
   	\sigma_g = \frac{G^2 m_a^2}{\delta v^4} \,,
   \end{equation}
   where $\delta v$ is the spread of speed of the axions, we note that it only diverges as a result of the forward scatting pole $\delta v \rightarrow 0$. It is not clear how this leads to a physical effect. Moreover, if any effect due to gravity exists, it is not evident to us why it should only occur for axions and not also for other potential forms of  light dark matter.}
For us, it is only important that whatever the effect is, it can be described as classical gravitational self-interaction. A quantum treatment is not necessary. In particular, there is no reason why classical simulations of dark matter evolution should fail. 

Finally, we want to make a brief remark about the classicality of coherent states. In \cite{sikivieCoherent}, it is claimed that coherent states fail to reproduce a classical evolution even when their occupation number $N$ is infinite. We want to point out that this observation is only an artifact of an unphysical limit. Namely, the authors of \cite{sikivieCoherent} take $N\rightarrow \infty$ while keeping the coupling $\alpha_a$ fixed. This does not correspond to the classical limit but to an infinite amplitude of oscillations: $A\rightarrow \infty$. When we write the quantum break-time due to $2\rightarrow2$-scattering \eqref{time2to2} as
\begin{equation}
	t_{\text{q}} = \bar{m}_a^{-1} \frac{1}{\alpha_a^2 N} \,,
	\label{time2to2aN}
\end{equation}
we see that the limit of infinite amplitude implies that $t_{\text{q}} \sim 1/N$. This is the scaling also observed in \cite{sikivieCoherent} (see figure 2 there).

   As is clear from equation \eqref{collectiveGeneral}, the limit $N\rightarrow \infty$ with fixed coupling $\alpha_a$ corresponds to an infinite collective coupling: $\lambda \rightarrow \infty$. 
In the language of many-body analysis \cite{nico}, this means that the  axion gas is {\it infinitely} overcritical.   Thus, what we observe is not surprising.
As we know from \cite{nico}, the quantum break-time can shorten significantly in the overcritical regime. This happens because in this regime, the attractive homogeneous Bose-gas is unstable, with Lyapunov exponent set by $\lambda$. Since $\lambda$ scales as $N$ in the above limit ($N\rightarrow \infty$ with fixed $\alpha_a$), the quantum break-time must scale as $1/N$.\footnote
{ See \cite{hertzbergClassical} for a suggestion to extend the validity of the classical solution in the overcritical regime by averaging over a set of random initial conditions.}

In any case, the instabilities developed  in the artificial limit $\lambda \rightarrow \infty$ are irrelevant for axion physics since as we have shown explicitly in section \ref{sec:collective}, this domain is not applicable for the realistic axion field, which is safely subcritical: $\lambda \ll 1$.
   
 As discussed in section \ref{sec:scaling}, one obtains the correct form of the classical limit by taking $\hbar\rightarrow0$. In this case, we have $N\rightarrow \infty$ while the collective coupling $\alpha_a N$ stays fixed. 
Since this implies that $\alpha_a \sim 1/N$, equation \eqref{time2to2aN} leads to the scaling $t_{\text{q}} \sim N$. Thus, the classical description stays valid indefinitely in the classical limit $N\rightarrow \infty$. 

\section{Conclusion}
The aim of the present paper is twofold. First, we wanted to determine the quantum break-time of the hypothetical dark matter axion field in our Universe, i.e., the time-scale during which it can be described classically. 
 The phenomenon of  quantum breaking for a coherently oscillating nonlinear scalar field was studied previously in \cite{us} and we have adopted a similar approach for axions. 
First, we represented the classical axion field as expectation value over an underlying quantum state. Subsequently, we calculated the rate of scattering among individual axions to estimate the quantum break-time as time-scale on which the expectation value over this state deviates from the classical solution. Equations \eqref{time4to2} and \eqref{time2to2} constitute our result, from which we concluded that the quantum break-time exceeds the age of the Universe by many orders of magnitude. Concerning experimental axion searches, we therefore have the clear-cut message: The approximation of the axion field as classical oscillating gas is extremely accurate and safe for all practical purposes. 

The second goal of our paper was to study quantum breaking in general. First, we discussed in section \ref{sec:scaling} that a generic quantum system possesses two fundamentally different couplings: the quantum coupling $\alpha$, which describes the interaction between individual quanta and vanishes in the classical limit $\hbar \rightarrow 0$, and the collective coupling $\lambda$, which parameterizes the strength of classical nonlinearites and is independent of $\hbar$.  The collective coupling  is a property of a given state as it depends on the occupation number of interacting quanta.
Since the quantum break-time becomes infinite for $\hbar \rightarrow 0$ in any system, it  cannot solely depend on the collective coupling $\lambda$. Instead, it must necessarily also have a dependence on the quantum coupling $\alpha$. From this reasoning, it became apparent in section \ref{sec:otherWork} that any short time-scale associated to axion dynamics originates from classical nonlinearities, which are controlled by  $\lambda$, and is not related to quantum breaking. In section \ref{sec:collective},  generalizing the results of  \cite{nico},  we additionally studied the possibility that a system can exhibit fast quantum breaking if it is overcritical, $\lambda > 1$, and exhibits a classical instability 
quantified by a non-zero Lyapunov exponent. For the case of the axion, we showed that neither condition is fulfilled so that fast quantum breaking cannot occur.  Our general conclusions equally apply to other light 
scalar or pseudo-scalar candidates of dark matter that are treated as 
classical fields.  

 \section*{Acknowledgements}
 We thank C\'{e}sar G\'{o}mez and Georg Raffelt for useful discussions. The work of G.D. was supported by the Humboldt Foundation under Alexander von Humboldt Professorship, the ERC Advanced Grant  "Selfcompletion"  (Grant No. 339169), FPA 2009-07908, CPAN (CSD2007-00042), HEPHACOSP-ESP00346, and by TR 33 "The Dark Universe".

\end{document}